\documentclass[twoside]{article}
\usepackage{fleqn,espcrc2}
\usepackage{graphicx}

\newcommand{\AmS}{{\protect\the\textfont2
  A\kern-.1667em\lower.5ex\hbox{M}\kern-.125emS}}

\title{A Rational Approach to the Resonance Region}

\author{Pere~Masjuan~\address{
        Grup de F{\'\i}sica Te{\`o}rica and IFAE,
        Universitat Aut{\`o}noma de Barcelona, 08193 Barcelona, Spain.}
        \thanks{Talk given at the 14th International QCD Conference, QCD 08,
        7-12 July 2008, Montpellier (France). This work is based
        on Ref.~\cite{PerisMasjuan07}.
I would like to thank the organizers for the nice atmosphere
during the conference.
This work has been supported by
CICYT-FEDER-FPA2005-02211, SGR2005-00916, the Spanish Consolider-Ingenio 2010
Program CPAN (CSD2007-00042)
and by the EU Contract No. MRTN-CT-2006-035482, ``FLAVIAnet''.
}
        }

\begin{document}

\begin{abstract}
Resonance Saturation in QCD can be understood in the large-$N_c$ limit from the mathematical theory of Pad\'{e} Approximants to meromorphic functions.

\end{abstract}

\maketitle

\section{Introduction}

The Chiral Lagrangian organizes the physics of the strong interactions at low energy as an expansion in powers of momentum and masses of the lightest pseudoscalar fields. Since all the heavier states of QCD are integrated out, their physics is encoded in a set of Low Energy Constants (LECs). These LECs are indispensable to make definite predictions in Chiral Perturbation Theory. However, while at $\mathcal{O}(p^4)$ the LECs are relatively well known, at $\mathcal{O}(p^6)$, most of the $O(100)$ LECs are completely unknown.

At the same time, QCD Green's functions seem to be approximately saturated by a few resonances. In the vector channel, that fact was called Vector Meson Dominance. Also, LECs seem to be well saturated by the lowest meson (after constraints in the OPE) \cite{Donoghue89}.

The theoretical support for these phenomenological ideas comes out from the framework of Large-$N_c$ QCD. The main point is that in Large-$N_c$ QCD mesons are stable \cite{thoftLN}. With this consideration and keeping only a finite set of resonances (instead of an infinite number of mesons in Green's functions),\cite{PerisMHA} proposed the Minimal Hadronic Approximation ($MHA$). The question now is that MHA only uses a finite set of resonances. How the error can be estimated and how the approximation can be improved is the main goal of \cite{PerisMasjuan07} within the mathematical theory of Pad\'{e} Approximants.

Let $f(z)$ be a function of a complex $z$ with and expansion $f(z)=\sum_{n=0}^{\infty}f_n z^n$ when $z\rightarrow 0$. Then, we can define a rational function $P_N^M(z)$ such that $P_N^M(z)\equiv \frac{Q_M(z)}{R_N(z)}\approx f_0 + f_1 z + f_2 z^2 + \cdots + f_{M+N} z^{M+N} + \mathcal{O}(z^{M+N+1})$. Then, $P_N^M(z)$ is a Pad\'{e} Approximant (PA).

The convergence properties of the PAs to a given function are much more difficult than those of normal power series.  However, those which concern meromorphic functions are rather well known. The main result which we will use is Pommerenke's Theorem \cite{Pommerenke} which asserts that the sequence of PAs to a meromorphic function is convergent everywhere in any compact set of the complex plane except, perhaps, in a set of zero area. This set includes the set of poles where the original function $f(z)$ is ill-defined but there may include extraneous poles as well. The previous convergence theorem requires that, either these extraneous poles move very far away from the region as the order of the PA increases, or they pair up with a close-by zero (Fig.~\ref{PA50}) becoming what is called a \textit{defect} \cite{Baker}. These must be considered artifacts of the approximation. Near these extraneous poles the approximation breaks down, but away from them, the approximation is safe.

In the physical case, the original function $f(z)$ will be a Green's function of the momentum variable $Q^2$. In Large-$N_c$ QCD this Green's function is meromorphic with all its poles located on the negative real axis in the complex $Q^2$ plane. Its poles are identified with the meson masses. On the other hand, the region to be approximated by the PAs will be that of euclidean values for the momentum $Q^2>0$. The expansion of the Green's function for large and positive $Q^2$ coincides with the Operator Product Expansion.
We will see that some of the poles and residues of the PAs may become complex. The reason lies in the fact that a meromorphic function does not obey any positivity constraints. This means that these poles and residues may have nothing to do with the physical meson masses and decay constants. However, and this is pointed out in \cite{PerisMasjuan07}, this does not spoil the validity of the rational approximation provided the poles, complex or not, are not in the region of $Q^2$ one is interested in. It is to be considered rather as the price to pay for using a rational function, which has only a finite number of poles, as an approximation to a meromorphic function with an infinite set of poles.

\section{The model}

To exemplify these characteristics, let consider the two-point functions of vector and axial-vector currents in the chiral limit
\begin{eqnarray} \label{correlator}
\Pi^{V,A}_{\mu\nu}(q)=\ i\,\int d^4x\,e^{iqx}\langle J^{V,A}_{\mu}(x)
J^{\dag\ V,A}_{\nu}(0)\rangle = \\\nonumber
\left(q_{\mu} q_{\nu} - g_{\mu\nu} q^2 \right)\Pi_{V,A}(q^2) \ ,
\end{eqnarray}

with $J_{V}^\mu(x) = {\overline d}(x)\gamma^\mu u(x)$ and $J_A^\mu(x) = {\overline
d}(x)\gamma^\mu \gamma^5 u(x)$.
Following Refs. \cite{Shifman}, we define our model as ($\Pi_{LR}(q^2)=\frac{1}{2}(\Pi_V(q^2)-\Pi_A(q^2))$):


\vspace*{-0.5cm}

\begin{eqnarray}\label{oneprime}
&&\Pi_{LR}(q^2)= \frac{F^2_0}{q^2}+ \frac{F_{\rho}^2}{-q^2+M_{\rho}^2}\\ \nonumber
&+&  \sum_{n=0}^{\infty} \left\{\frac{F^2}{-q^2+M^2_V(n)}-
    \frac{F^2}{-q^2+M^2_A(n)}\right\}
\end{eqnarray}

Here $F_\rho,M_\rho $ are  the electromagnetic decay constant
and mass of the $\rho$ meson and $F_{V,A}(n)$ are the electromagnetic decay constants of the $n-th$
resonance in the vector (resp. axial) channels, while $M_{V,A}(n)$ are the corresponding masses.
$F_0$ is the pion decay constant in the chiral limit. The dependence on the resonance excitation
number $n$ is $F^2_{V,A}(n)=F^2= \mathrm{constant}$ and $M_{V,A}^2(n) = m_{V,A}^2 + n \ \Lambda^2$
in accord with known properties of the large-$N_c$ limit of QCD as well as alleged
properties of the associated Regge theory \cite{thoftLN}.

To resemble the case of QCD, we will demand that the usual parton-model logarithm is reproduced in
both vector and axial-vector channels and that $\Pi_{LR}(q^2)$ has an operator
product expansion which starts at dimension six. A set of parameters satisfying these conditions, while keeping the model realistic and at a manageable level, is given by
$F_{0}= 85.8 \,\,{\mathrm{MeV}}$, $F_{\rho}=133.884 \,\,{\mathrm{MeV}}$, $F=143.758 \,\,{\mathrm{MeV}}$, $M_{\rho}=0.767 \,\,{\mathrm{GeV}}$, $m_A=1.182 \,\,{\mathrm{GeV}}$, $m_V=1.49371
\,\,{\mathrm{GeV}}$ and $\Lambda=1.2774 \,\, {\mathrm{GeV}}$.
The values
are chosen so that:
\begin{equation}\label{chiralexp}
Q^2 \Pi_{LR}(Q^2)|_{Q^2\rightarrow 0}\approx C_0 + C_2 Q^2 + C_4 Q^4 + \cdots
\end{equation}
\vspace{-0.25cm}

\begin{equation}\label{OPEexp}
Q^2 \Pi_{LR}(Q^2)|_{Q^2\rightarrow \infty }\approx 0 + \frac{0}{Q^2} + \frac{C_{-4}}{Q^4} + \frac{C_{-6}}{Q^6}+ \cdots
\end{equation}
\vspace{-0.3cm}
\subsection{PA with a Model}

The simplest PA to the function $Q^2 \Pi_{LR}(- Q^2)$ with the right fall-off as $Q^{-4}$ at large $Q^2$ is $P_2^0 (Q^2)$, which has three unknown coefficients fixed with the first three coefficients from the chiral expansion of (\ref{chiralexp}), i.e., $C_{0,2,4}$:
\vspace{-0.25cm}
\begin{eqnarray}\label{P02}
    P^{0}_{2}(Q^2)=\frac{-\ r_R^2}{(Q^2+z_R) (Q^2+z_R^*)}
\end{eqnarray}

where $r_R^2=3.379\times 10^{-3}$ and $z_R=0.6550+i\ 0.1732$.


\begin{figure}[htb]
   \includegraphics[width=6.8cm]{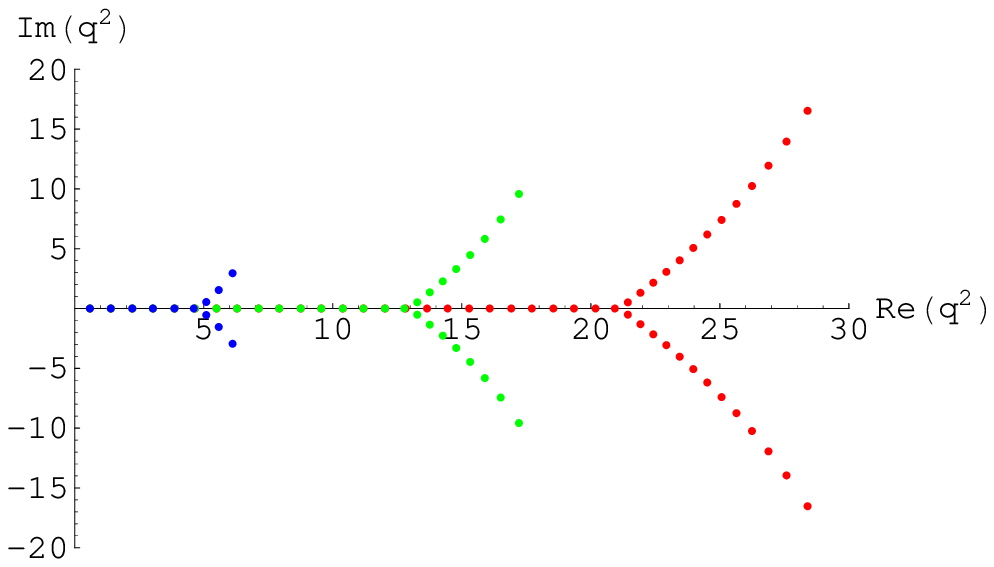}\\
  \includegraphics[width=6.8cm]{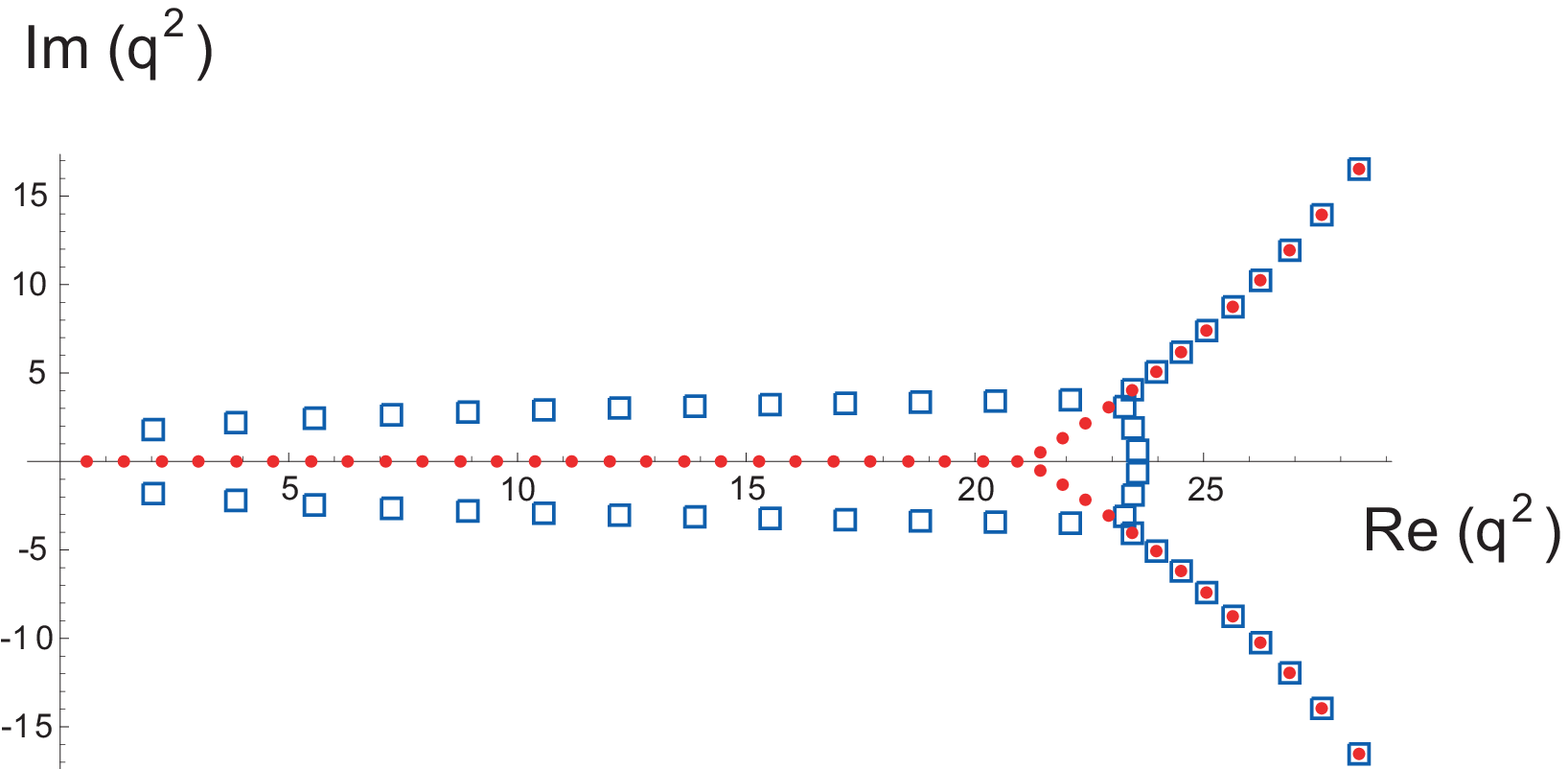}\\
  \vspace{-0.5cm}
  \caption{{\small {\bf Top}: Location of the poles of $P_{12}^{10}$ (green), $P_{32}^{30}$ (blue) and $P_{52}^{50}$ (red) in the complex $q^2$ plane. When the order of the PA is increased, the overall shape of the figure does not change but the two branches of complex poles move toward the right, away from the origin; {\bf Bottom}: Location of the poles (dots) and zeros (squares) of $P_{52}^{50}$ in the complex $q^2$ plane.}}\label{PA50}
\end{figure}

A pair of complex-conjugate poles means that these poles cannot be interpreted in any way as the meson states appearing in the physical spectrum. In spite of this, if one expands (\ref{P02}) for large $Q^2>0$, one finds $C_{-4}=-r^2_R=3.379\times 10^{-3}$ which differs from the real value by about $30\%$. Even better is the prediction of the fourth therm in the chiral expansion ($3\%$ of error). This agreement is not a numerical coincidence and the approximation can be systematically improved if more terms of the chiral expansion are known (Fig.~\ref{PA50}-top). For example, going up to $P_{52}^{50}$ (with $103$ parameters), one finds that this rational approximant correctly determines the values for $C_{-4,-6,-8}$ with $52,48,45$ decimal figures respectively. In the case of $C_{206}$, which is the first predictable term from the chiral expansion, the accuracy reaches $192$ decimal figures.

As it happens for (\ref{P02}), also higher order PAs may develop some artificial poles. Fig.~\ref{PA50}-bottom shows the location of the $52$ poles of the PA $P_{52}^{50}$ in the complex $q^2$ plane. A detailed numerical analysis reveals that the poles and residues reproduce very well the
value of the meson masses and decay constants for the lowest part of the physical spectrum of the
model given in (\ref{oneprime}), but the agreement deteriorates very quickly as one
gets farther away from the origin, eventually becoming the complex numbers seen in Fig.
\ref{PA50}. It is by creating these analytic defects that rational functions can effectively mimic
with a finite number of poles the infinite tower of poles present in the original function
(\ref{oneprime}).
Then, the determination of decay constants and masses extracted as the residues and
poles of a PA deteriorate very quickly as one moves away from the origin. There is no reason why
the last poles and residues in the PA are to be anywhere near their physical counterparts and their
identification with the particle's mass and decay constant should be considered unreliable.
Clearly, this particularly affects low-order PAs.

\begin{table*}[!t]
\setlength{\tabcolsep}{1pc}
\centering
\catcode`?=\active \def?{\kern\digitwidth}
\begin{tabular*}{\textwidth}{@{} l@{}ccccc|ccc}
  \hline
 & $C_0$ & $C_2$ & $C_{4}$ & $C_6$ & $C_8$ & $C_{-4}$ & $C_{-6}$ & $C_{-8}$  \\
  \hline
 & $-F_0^2$ & $-4\,L_{10}$ & $-43\pm 13$ & $81\pm 53$ & $-145\pm 120$ & $-4.1\pm 0.5$
 & $6\pm 2 $& $-7\pm 6$   \\
  & &  & $ -8\textit{C}_{87}$ &  &  &  &   &   \\
  \hline
\end{tabular*}
\caption{{\small\emph{Values of the coefficients $C_{2k}$ in the high- and low-$Q^2$ expansions of $Q^2\
\Pi_{LR}(-Q^2)$ in Eq. (\ref{chiralexp},\ref{OPEexp}) in units of $10^{-3}\ GeV^{2-2k}$. Recall that $C_{-2}=0$.
}}}\label{table2}
\end{table*}

\section{A QCD case}

In the case of large-$N_c$ QCD in the chiral limit, two considerations arise. First, the lack of input data makes going to higher order in the construction of the rational approximants a difficult work. In addition, an order of PAs is increased by one unit by adding two new inputs. Second, any input value will have an error (both experimental and because of chiral and large-$N_c$ limits), and this error will propagate through the rational approximant. In spite of these difficulties one may feel encouraged by the phenomenological fact that resonance saturation approximates meson physics rather well.

The simplest PA to the function $Q^2 \Pi_{LR}(- Q^2)$ with the right fall-off as $Q^{-4}$ at large $Q^2$ is $P_2^0 (Q^2)=\frac{a}{1+ A \ Q^2 + B\ Q^4}$. The values of the three unknowns $a, A, B$ may be fixed by requiring that his PA reproduces the correct values for $F_0, L_{10}$ and $\delta M_{\pi}$, the electromagnetic pion mass difference (which is defined in terms of $\Pi_{LR}$ as $\delta M_{\pi}^2 \equiv M_{\pi^+}^2- M_{\pi^0}^2 = \ -\frac{3}{4 \pi} \frac{\alpha}{f_0^2} \int_0^{\infty} dQ^2\ Q^2 \Pi_{LR}(Q^2)$). Then, we use: $F_0= 0.086\pm  0.001\ \mathrm{GeV}$, $\delta M_{\pi}= 4.5936\pm 0.0005\ \mathrm{MeV}$, $L_{10}(0.5\ \mathrm{GeV}) \leq L_{10}\leq L_{10}(1.1\ \mathrm{GeV}) \Longrightarrow $, $L_{10}=\left( -5.13\pm 0.6\right) \times 10^{-3}$.

The final results are shown in table \ref{table2}. In a real QCD case, the lack of input data does not allow us systematically improve our results. However, other Pad\'{e} Approximants are useful for doing this improvement, for instance, what is called Pad\'{e}-Type Approximants (PT). PTs are defined as the ratio $T^M_N$ of two polynomials where the polynomial in the denominator has its zeros given in advance, preassigned, for example, at the position of the first $N$ resonance masses on the original Green's function. Therefore, we will use for the masses \cite{PDG}: $m_{\rho}= 0.7759 \pm 0.0005$, $m_{\rho'}=1.459 \pm 0.011$, $m_{\rho''}=1.720 \pm 0.020$, $m_{\rho'''}= 1.880 \pm 0.030$,$m_{a_1}=1.230 \pm 0.040$ and $m_{a'_1}= 1.647 \pm 0.022$ (all numbers in GeV). Because we can compute succesive terms in a PT sequence, we can get a feeling of convergence and estimate in this way the error of our approximation. Our final result is $C_{87}=(5.7 \pm 0.5)\times 10^{-3}GeV^{-2}$. Recently, \cite{JJRosellPichC87} has found $C_{87}({\small \mu=0.77GeV})=(3.9\pm 1.4)\times 10^{-3}GeV^{-2}$ at NLO in $1/N_c$ in Resonance Chiral Theory.

\vspace{-0.2cm}

\begin{figure}[htb]
  \includegraphics[width=6.8cm]{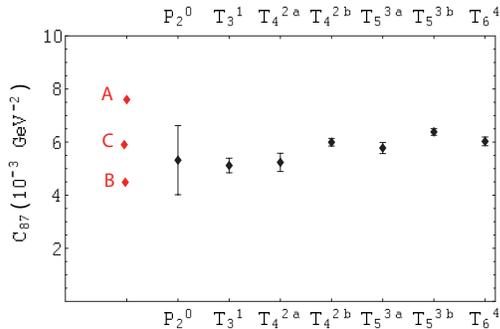}
    \vspace{-0.5cm}
  \caption{{\small Prediction for $C_{87}$ in the large-$N_c$ limit from  the PA $P^0_2$ and PTs in Ref. \cite{PerisMasjuan07}. For comparison we also show the estimate from Refs. \cite{ABT},
   which we label `A', `B' and `C' (resp.).}}\label{fig:C87}
\end{figure}

\vspace{-0.3cm}
\section{Conclusion}

Ref. \cite{PerisMasjuan07} pointed out that approximating Large-$N_c$ QCD with a finite number of resonances may be reinterpreted within the mathematical Theory of Pad\'{e} Approximants to meromorphic functions. With the help of a toy model, we have reviewed the main results of this theory: rational approximants to a meromorphic Green's functions yield an accurate description of the Green's function in the Euclidean region $Re(q^2)<0$ while creating artificial poles (and corresponding residues) in the Minkowsi region $Re(q^2)>0$. Therefore, althoug they may be a very good approximation in the euclidean, it is in general unreliable to extract properties of individual mesons, such as masses and decay constants from an approximation to Large-$N_c$ QCD with only a finite number of states.

Pad\'{e} Approximants beyond Large-$N_c$ QCD have been applied, recently, to the study of the pion form factor \cite{JJPerisMasjuan}.

\end{document}